\documentclass[10pt,conference]{IEEEtran}
\IEEEoverridecommandlockouts

\usepackage{subfigure}
\usepackage{tcolorbox}
\usepackage{xspace}
\usepackage{enumitem}
\usepackage{bbding}
\usepackage{url}
\usepackage{booktabs}
\usepackage{tabularx}
\usepackage{circledsteps}
\usepackage{graphicx}
\usepackage{footmisc}
\usepackage{adjustbox}
\usepackage{pifont}
\usepackage{svg}
\usepackage{tikz}
\usepackage{xcolor}
\usepackage{listings}
\usepackage{cite}
\usepackage{float}

\usepackage{pifont}

\usepackage{multicol}

\usepackage{makecell}
\usepackage{multirow}
\usepackage{colortbl}
\newcolumntype{C}[1]{>{\centering\arraybackslash}p{#1}}

\usepackage{amsmath,bm,amsthm}
\theoremstyle{definition}

\usepackage{algorithm}
\usepackage{algpseudocode}
\algnewcommand\algorithmicforeach{\textbf{for each}}
\algdef{SE}[FUNCTION]{Function}{EndFunction}[2]{\algorithmicfunction\ \textnormal{#1}\ (#2)}{\algorithmicend\ \algorithmicfunction}%
\algdef{S}[FOR]{ForEach}[1]{\algorithmicforeach\ #1\ \algorithmicdo}
\algrenewcommand\alglinenumber[1]{\footnotesize #1}
\algrenewcommand\algorithmicrequire{\small \textbf{input:}}
\algrenewcommand\algorithmicensure{\small \textbf{output:}}
\algrenewcommand\algorithmicfunction{\textbf{Function}}
\algtext*{EndFunction}
\algtext*{EndIf}
\algtext*{EndFor}
\algtext*{EndWhile}

\usepackage{fontawesome}
\usepackage{xcolor}
\usepackage{framed}
\usepackage{textcomp}
\usepackage[capitalise]{cleveref}

\usepackage{microtype}
\usepackage{ifthen}

\newcommand{\code}[1]{\texttt{\footnotesize #1}}
\newcommand{\smalltitle}[1]{\noindent\textbf{#1}}

\definecolor{summarybg}{RGB}{240,240,240}

\newcounter{rq}

\begin{document}

\title{Understanding Agent-Reactive Bugs at the Model-Harness Boundary: An Empirical Study of LLM Agent Issue Reports}

\author{
\IEEEauthorblockN{Jingyi Chen, Songqiang Chen, Hengcheng Zhu, Jialun Cao\textsuperscript{*}, Jiasi Shen\textsuperscript{*}, and Shing-Chi Cheung}
\IEEEauthorblockA{The Hong Kong University of Science and Technology, Hong Kong, China\\
Guangzhou HKUST Fok Ying Tung Research Institute, Guangzhou, China\\
\{jchenix,i9s.chen,hzhuaq\}@connect.ust.hk\\
\{jcaoap,sjs,scc\}@cse.ust.hk}
\thanks{\textsuperscript{*}Corresponding authors: Jialun Cao and Jiasi Shen.}
}

\maketitle

\begin{abstract}
LLM agents span command-line interfaces (e.g., Codex) and agent frameworks (e.g., LangChain), integrating backend LLMs with harness code that parses model outputs, controls agent loops, and manages context. Both the harness and LLM-generated responses jointly shape an agent's execution. This architecture gives rise to bugs that cannot be readily understood by inspecting either component alone, because some bugs occur only when a particular LLM response elicits an abnormal reaction from the agent. Prior empirical studies of agent bugs have largely attributed failures either to limited model capabilities or to harness-side defects, such as outdated APIs and configuration misalignment, without characterizing these AR bugs. We conduct the first empirical study focused on agent-reactive (AR) bugs. Through manual analysis of 255 bug reports from Codex, Gemini-CLI, LangChain, and CrewAI, we construct a two-axis taxonomy covering observable symptoms and the LLM behaviors that trigger them. Our findings show that many AR bugs manifest as silent errors without well-defined test oracles, which makes detection difficult. The stochasticity of LLM responses further complicates bug reproduction. We additionally examine fixes proposed by users and implemented by developers. This analysis exposes a mismatch: users frequently advocate harness-side guardrails, whereas developers may attribute the issue to the LLM or respond slowly to user-proposed fixes. These findings point to the need for mechanisms that help users and developers understand the root causes and resolutions of AR bugs. Overall, the study highlights challenges specific to LLM agents and motivates the design of test oracles, reproduction support, and fault-localization techniques for AR bugs.
\end{abstract}

\begin{IEEEkeywords}
	LLM Agent Reliability, Model-Harness Interaction, Tool-Use Failure, Empirical Study
\end{IEEEkeywords}

\section{Introduction}

Large Language Model (LLM) agents, available in a form ranging from command-line interfaces (CLIs) to programmable agent frameworks, are increasingly being adopted for coding tasks~\cite{jimenez2024swebench,yang2024sweagent}, web navigation~\cite{zhou2024webarena}, and customer service~\cite{yao2024taubench}.
In the rest of this paper, we use \emph{agents} to refer to LLM agents for brevity.
At the core of an agent is a harness scripted to dispatch tool calls, parse model output, execute the agent loop, and manage context, while its backend LLMs are continually prompted to generate answers that can vary across runs. The agent reacts to these answers and performs the assigned task~\cite{ouyang2023nondeterminism}.
The highly reactive nature of interactions between the harness and the underlying stochastic LLMs distinguishes agents from traditional software. 
Evolving quickly, these agents also suffer from various bugs. Users of open-source agents often report the encountered bugs in the agent projects' GitHub trackers.

The bug reports of agents have drawn growing research attention.
Many empirical studies have been conducted to propose a taxonomy of agent bugs by mining issues from the repository of agents or agent frameworks~\cite{islam2026agents,zhu2026bugs,shah2026characterizing,zhang2026dissecting}. Studies have also been carried out to evaluate agent capabilities using various benchmarks~\cite{jimenez2024swebench,liu2024agentbench,yao2024taubench,mialon2024gaia} and explain incorrect LLM responses, such as hallucination,~\cite{huang2023hallucination,kokane2024toolscan,qin2024toolllm,xu2024toolhallucination,zhou2023ifeval,skripko2025iffunctioncalling,liu2024lostmiddle}.
These studies help explain the kinds of failure that can occur in agents and the limits of their capabilities. However, they have not examined a unique class of bugs that arise from LLM-harness interactions when the harness does not handle that LLM behavior properly, which we name as \textit{\textbf{agent-reactive (AR) bugs}}. 

Our analysis reveals that AR bugs account for 8.4\% of all actively discussed issues in four popular agent projects: Codex, Gemini-CLI, LangChain, and CrewAI (Section~\ref{sec:datacollection}). 
\begin{figure}[t]
\centering
\includegraphics[width=.94\linewidth]{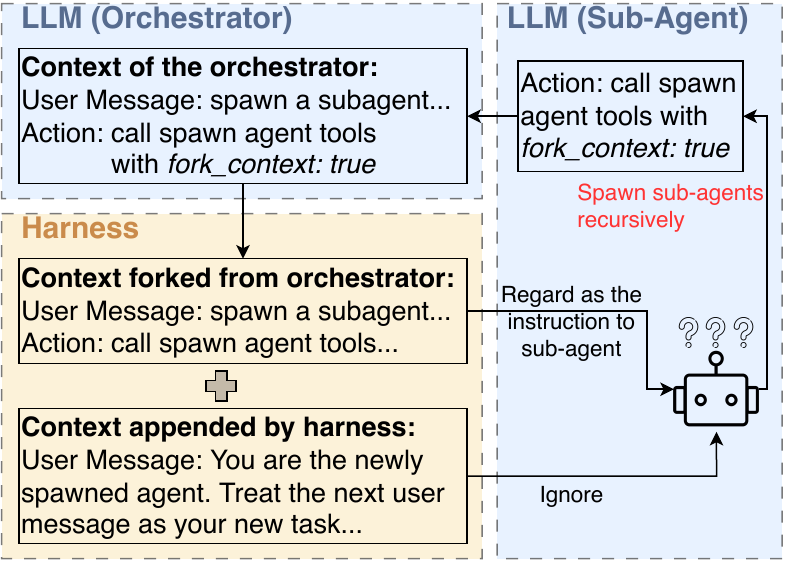}
\caption{Codex issue 13491: a forked worker inherits parent intent and attempts recursive delegation instead of following its assigned task.}
\label{fig:codex13491}
\end{figure}
 
For example, in Codex issue 13491 (shown in \cref{fig:codex13491}), the orchestrator model is requested to spawn a sub-agent with \texttt{fork\_context=true} setup. The harness forks the context from the orchestrator and appends handoff instructions informing the sub-agent to treat the next user message as its own task. However, the subagent model ignores the handoff instructions, acts as another orchestrator, and spawns sub-agents recursively.
Such AR bugs pose particular challenges to diagnose compared with both LLM-side performance problems and harness-side defects, like outdated APIs and configuration misalignment, because AR bugs cannot be identified by examining the harness or the LLM alone. The stochastic nature of LLMs also makes AR bugs hard to reproduce. Thus, understanding the symptoms and fixing strategies of the AR bugs is crucial to improving the reliability of agents.

To fill the gap, we conduct the first empirical study of AR bugs.
We collect agent bugs from two popular agent CLIs (Codex and Gemini-CLI) and two agent frameworks (LangChain and CrewAI). These four projects allow users to report agent failures as GitHub issues, where reporters, other users, and developers discuss symptoms, reproduction attempts, workarounds, and possible fixes. Based on the issues of the four projects, we obtain 255 AR bugs after rule-based filtering to obtain bug reports with LLM invocation and manual annotation to collect bugs that depend on specific LLM responses. 
We analyze the symptoms observed by users (RQ1) and the LLM behaviors that trigger them (RQ2), which motivates the design of test oracles and reproduction techniques. We study a follow-up question: when did these failures surface, and where were their fixes applied---in the agent harness or at the backend LLM? We answer this question by analyzing user proposals and developer actions in issue discussions and related linked pull requests (RQ3).

We identified five categories of AR bug symptoms and eight LLM behaviors that trigger AR bugs. Our findings reveal several challenges for detection, reproduction, and fix localization. 
First, many symptoms lack a rigorous oracle~\cite{barr2015oracle,tambon2024silent}, especially silent errors, where a plausible but incorrect result is hard to tell from a correct one. This suggests that agent users should carefully monitor and audit the LLM-harness interactions on critical tasks, as well as motivates research opportunities for designing effective test oracles to automatically detect AR bugs.
Second, we found that the occurrence of AR bugs is affected by specific LLM responses, context length, and the state of the workspace~\cite{ouyang2023nondeterminism,luo2014flaky}, making them difficult to reproduce. Thus, developers are recommended to instruct users to properly record and share the context to ease debugging of AR bugs. This also leaves auto-reproduction of AR bugs a significant research problem.
Third, we reveal that the diagnosis of AR bugs can be controversial: users may propose harness-side guardrails, while developers may attribute failure to poor LLM capability. This poses the need for a guideline for fault attribution in agents, which can help developers and users align on the root cause of AR bugs and improve the reliability of agents.

In summary, our study makes four contributions:
\begin{itemize}
    \item We conduct the first empirical study dedicated to agent reactive (AR) bugs, whose manifestation depends on both a specific LLM behavior and the harness that handles it. 
    \item We manually annotate a dataset of 255 AR bugs from two agent CLIs (Codex and Gemini-CLI) and two agent frameworks (LangChain and CrewAI). 
    \item We build a two-axis taxonomy that pairs the symptoms of these bugs with the LLM behaviors that trigger them, and use it to analyze how a behavior turns into a symptom. We find the new challenges raised from AR bugs in detection, reproduction, and fix localization, motivating new research opportunities.
    \item We study how users and developers currently act to fix these bugs, revealing a mismatch between them over fault attribution in agents and inspiring how to handle users' complaints in the future.
\end{itemize}

\section{Preliminaries}
\label{sec:preliminaries}
\subsection{LLM Agents}
\label{sec:bg-agents}

Large language models (LLMs) have shown strong capability in software engineering~\cite{hou2024llmse}, including code generation~\cite{chen2021codex,chen2025apidoc}, code translation~\cite{chen2026pseudocode}, and test generation~\cite{schafer2024unittest}. To move beyond answering a single prompt, LLMs are wrapped in an \emph{agent loop} and connected to external tools, yielding an LLM agent that can plan and carry out multi-step tasks~\cite{xi2023rise,yang2024sweagent}. In common designs, the loop alternates between behaviors generated by the model and results returned by tools or environments~\cite{yao2023react,yang2024sweagent}. Such agents are deployed as clients (e.g., OpenAI Codex and Gemini-CLI) and frameworks for customized agents (e.g., LangChain and CrewAI).
Following prior work~\cite{yang2024sweagent,xi2023rise,meng2026harness}, we view an agent as two interacting components. 

\textbf{Backend LLM.} Given the current context, the backend LLM generates the next response or action. Its output may vary across runs even with an identical initial prompt~\cite{ouyang2023nondeterminism}. 

\textbf{Harness.} It provides an agentic wrapping of the backend LLM~\cite{meng2026harness}. It renders prompts, parses model output, performs the agent loop, manages context, and so on. The formats or constraints that the LLM should follow and that the harness is expected to handle are defined in a piece of harness code. 

The harness reacts to the LLM. It sends the LLM a context with task instructions and previous responses. The LLM returns either a final response or an action, such as a tool invocation. The harness parses the output, performs the action if needed, appends the result to the context, and continues until the task ends. Unlike traditional software, whose control flow is predefined, an agent's control flow is determined on-the-fly by the interaction between the harness and the LLM. For example, in the agent shown in \cref{fig:codex13491}, the LLM in the sub-agent treated the forked orchestration history as an instruction for itself and invoked the spawn-agent tool. The control flow of the whole system therefore shifts from the intended sub-agent task to a recursive delegation loop.

\subsection{Agent Bugs}
\label{sec:bg-bugs}
Users of agents routinely encounter bugs and report them on project GitHub trackers, sometimes with reproduction steps, workarounds, or proposed fixes. Many empirical studies analyze these reports through taxonomies of bug types, symptoms, and root causes~\cite{islam2026agents,zhu2026bugs,shah2026characterizing,zhang2026dissecting}. These studies often attribute a failure to one of the two components: limited backend-LLM capability, such as hallucination~\cite{huang2023hallucination,xu2024toolhallucination}, or harness-side defects, such as outdated APIs and configuration misalignment~\cite{zhu2026bugs,shah2026characterizing,zhang2026dissecting}.

In this work, we focus on bugs at the interface between the two components, which we call \emph{AR bugs}. Such bugs manifest only when the backend LLM exhibits particular behaviors, such as generating unexpected tool arguments, ignoring task instructions, or fabricating claims, and the harness handles that behavior in a way that produces a user-visible failure. These bugs can be hard to detect when their symptoms are inaccurate outputs rather than thrown exceptions. They can be hard to reproduce because the triggering behavior is generated by the LLM and may depend on long context, workspace state, and even model version. Whether the bugs should be fixed in the harness or in the backend LLM is often controversial.

\section{Data Collection}
\label{sec:datacollection}

To study AR bugs in agents, we mine the issues of four widely used open-source agent projects. Through automated filtering and manual annotation, we identify the subset of bug reports whose manifestation depends on a specific response from the backend LLM. The pipeline is summarised in \cref{tab:datacollection}.

\subsection{Subject Collection}
\label{ssec:dc:subjects}
We select four projects that cover two common forms of agents: agent CLIs and agent-construction frameworks. Each selected project is open source, has a public GitHub issue tracker, contains enough issue discussion for manual analysis, and implements or supports agent workflows with model calls, tool use, and so on. The two clients, Codex (\code{openai/codex}) and Gemini-CLI (\code{google/gemini-cli}), are applications that users can run directly. The two frameworks, LangChain (\code{langchain-ai/langchain}) and CrewAI (\code{crewai/crewai}), provide programmable abstractions for users to construct customized agents. This split lets us examine whether the symptoms and triggering LLM behaviors of agent bugs differ between agent CLIs and agent-construction frameworks. For each project, we collect all issues opened from the start of its issue tracker until April 2026.

\subsection{Issue Mining and Annotation}
\label{ssec:dc:pipeline}
For each subject project, we collect all its issues through the GitHub REST API, yielding a starting corpus of 32{,}373 issues across the four projects (\cref{tab:datacollection}, column \emph{Raw}).

We then use keyword-based filtering to discard non-bug issues (e.g., feature requests, questions, documentation suggestions) and bugs whose reproduction cannot reach an LLM call (e.g., missing-module errors, package-installation failures), reducing the corpus to 18{,}161 bug issues (\cref{tab:datacollection}, column \emph{Bugs}) and then to 11{,}288 LLM-reachable bug issues (column \emph{LLM-reachable}). 
We treat a bug as LLM-reachable if the reported reproduction steps contain the invocation of the backend LLM.
Finally, we keep only issues with sufficient maintainer or community engagement so that annotators have enough context to analyse. The four projects differ in user mix and discussion density: Codex and Gemini-CLI receive many user reports and comments but rarely have linked PRs, while LangChain and CrewAI attract fewer but technically deeper discussions. We therefore adapt the engagement threshold per project. Codex and Gemini-CLI require a linked PR \emph{or} at least three non-author commenters; LangChain requires a linked PR \emph{and} at least three non-author commenters; CrewAI requires a linked PR \emph{and} at least one non-author commenter, relaxed due to its smaller corpus. After all three filters, 3{,}037 issues (column \emph{Engaged}) are forwarded to manual annotation.

For every issue in the engagement-filtered pool, we ask two questions: \emph{(i)~does the bug's manifestation depend on a specific LLM response?} and, if so, \emph{(ii)~which LLM behavior triggers it and how does the bug manifest?} The first question yields a binary AR / non-AR label; the second yields the two-axis taxonomy (triggering behavior $\times$ symptom) used in \cref{sec:rq2}. Two authors annotated the issues following a two-phase protocol:

\smalltitle{Phase 1 — Initial taxonomy.}
Two authors independently sampled $5\%$ of the engagement-filtered pool and labeled the issues in an open-coding fashion, naming each triggering behavior and symptom category as it arose. If an issue does not depend on a specific LLM response (e.g., the issue is due to network or data streaming problems), we assign a special label for it. They then merged their category sets and refined the definitions through discussion to produce a shared initial taxonomy.

\smalltitle{Phase 2 — Annotation and reconciliation.}
Then the two authors partitioned the remaining $95\%$, and each labelled their portion against the shared taxonomy. New categories encountered during this phase were temporarily added and were merged and refined after the annotation. Conflicting or uncertain labels were then discussed until both annotators reached a consensus.

Overall, the two authors each spent about 150 hours on phase 1 and phase 2. Of the 3{,}037 engagement-filtered issues, 255 are AR bugs and 2{,}782 are non-AR bugs (\cref{tab:datacollection}, column \emph{AR bugs}).
For these 255 AR bugs, we further inspect issue discussions and linked pull requests to analyse user proposals and actions by project developers about where the fix should live; the detailed rubric is presented in \cref{sec:rq3}.

\begin{table}[t]
\caption{Issue counts at each stage of the data-collection pipeline. \emph{Raw}: issues crawled from GitHub. \emph{Bugs}: issues that report a bug. \emph{LLM-reachable}: bug issues whose reproduction steps can reach an LLM call. \emph{Engaged}: LLM-reachable bug issues with sufficient community engagement (see \cref{ssec:dc:pipeline}). \emph{AR bugs}: issues that human annotators confirmed as agent-reactive.}
\label{tab:datacollection}
\centering
\begin{adjustbox}{max width=\columnwidth}
\begin{tabular}{@{}lrrrrr@{}}
\toprule
\textbf{Project} & \textbf{Raw} & \textbf{Bugs} & \makecell{\textbf{LLM-}\\\textbf{reachable}} & \textbf{Engaged} & \makecell{\textbf{AR}\\\textbf{bugs}} \\
\midrule
Codex (CLI)        &  8{,}523 &  6{,}290 & 2{,}428 &    963 &  82 \\
Gemini-CLI         & 12{,}662 &  4{,}434 & 2{,}802 & 1{,}201 &  90 \\
LangChain          &  9{,}314 &  6{,}365 & 5{,}308 &    545 &  47 \\
CrewAI             &  1{,}874 &  1{,}072 &    750 &    328 &  36 \\
\midrule
\textbf{Total}     & \textbf{32{,}373} & \textbf{18{,}161} & \textbf{11{,}288} & \textbf{3{,}037} & \textbf{255} \\
\bottomrule
\end{tabular}
\end{adjustbox}
\end{table}

\section{Study Results and Analysis}
\label{sec:studyresults}
\subsection{RQ1: Symptoms}

To answer the first question, we analyze the \emph{symptoms}, i.e., what users observe when failures manifest, of each issue in our corpus. For each symptom, we give its definition, a representative issue, and its distribution across the four projects. We also summarize either users' attitudes or the difficulties they face for each symptom.

\providecommand{\rqTSZero}{\cellcolor[HTML]{F8FAFC}}
\providecommand{\rqTSCell}[2]{\cellcolor[HTML]{#1}#2}
\providecommand{\rqTSDark}[2]{\cellcolor[HTML]{#1}\textcolor{white}{#2}}
\providecommand{\rqTSTotal}[1]{\cellcolor[HTML]{EEF2F7}\textbf{#1}}
\providecommand{\rqTSGrandTotal}[1]{\cellcolor[HTML]{DBE5EF}\textbf{#1}}

\begin{table*}[t]
\centering
\caption{Triggering behaviors by symptom across projects. Each panel is one project: rows are triggering behaviors, columns are symptoms, darker cells indicate more issues, blank cells indicate zero, and the rightmost and bottom totals summarize triggering-behavior and symptom counts.}
\label{tab:trigger-symptom}
\setlength{\tabcolsep}{3.5pt}
\small
\begin{adjustbox}{max width=\textwidth}
\begin{tabular}{lrrrrrrlrrrrrrlrrrrrrlrrrrrr}
\toprule
\multicolumn{7}{c}{Codex} & & \multicolumn{6}{c}{Gemini-CLI} & & \multicolumn{6}{c}{LangChain} & & \multicolumn{6}{c}{CrewAI} \\
\cmidrule(lr){1-7}\cmidrule(lr){9-14}\cmidrule(lr){16-21}\cmidrule(lr){23-28}
Trigger & SE & Cr & EIO & RL & H & Total & & SE & Cr & EIO & RL & H & Total & & SE & Cr & EIO & RL & H & Total & & SE & Cr & EIO & RL & H & Total \\
\cmidrule(lr){1-7}\cmidrule(lr){9-14}\cmidrule(lr){16-21}\cmidrule(lr){23-28}
INC  & \rqTSDark{006D2C}{31} & \rqTSZero & \rqTSCell{DEF2F4}{1} & \rqTSCell{C8EAE6}{4} & \rqTSZero & \rqTSTotal{36} & & \rqTSDark{3EAB70}{19} & \rqTSCell{DEF2F4}{1} & \rqTSZero & \rqTSCell{CFEDEB}{3} & \rqTSCell{DEF2F4}{1} & \rqTSTotal{24} & & \rqTSCell{DEF2F4}{1} & \rqTSCell{DEF2F4}{1} & \rqTSCell{DEF2F4}{1} & \rqTSCell{DEF2F4}{1} & \rqTSZero & \rqTSTotal{4} & & \rqTSCell{CFEDEB}{3} & \rqTSZero & \rqTSZero & \rqTSZero & \rqTSZero & \rqTSTotal{3} \\
UTA  & \rqTSCell{C8EAE6}{4} & \rqTSZero & \rqTSCell{CFEDEB}{3} & \rqTSCell{B9E4DD}{6} & \rqTSCell{DEF2F4}{1} & \rqTSTotal{14} & & \rqTSCell{CFEDEB}{3} & \rqTSCell{B2E1D8}{7} & \rqTSCell{B9E4DD}{6} & \rqTSCell{7DCAAE}{13} & \rqTSZero & \rqTSTotal{29} & & \rqTSCell{DEF2F4}{1} & \rqTSCell{92D5C2}{11} & \rqTSZero & \rqTSZero & \rqTSZero & \rqTSTotal{12} & & \rqTSCell{DEF2F4}{1} & \rqTSCell{DEF2F4}{1} & \rqTSCell{C0E7E2}{5} & \rqTSZero & \rqTSZero & \rqTSTotal{7} \\
MTC  & \rqTSCell{B2E1D8}{7} & \rqTSZero & \rqTSZero & \rqTSZero & \rqTSZero & \rqTSTotal{7} & & \rqTSCell{DEF2F4}{1} & \rqTSZero & \rqTSCell{D6EFF0}{2} & \rqTSZero & \rqTSZero & \rqTSTotal{3} & & \rqTSCell{CFEDEB}{3} & \rqTSDark{48B07A}{18} & \rqTSCell{DEF2F4}{1} & \rqTSCell{DEF2F4}{1} & \rqTSZero & \rqTSTotal{23} & & \rqTSCell{AADFD4}{8} & \rqTSCell{CFEDEB}{3} & \rqTSCell{C8EAE6}{4} & \rqTSCell{DEF2F4}{1} & \rqTSZero & \rqTSTotal{16} \\
IHI  & \rqTSCell{DEF2F4}{1} & \rqTSZero & \rqTSCell{D6EFF0}{2} & \rqTSCell{DEF2F4}{1} & \rqTSCell{C0E7E2}{5} & \rqTSTotal{9} & & \rqTSZero & \rqTSCell{7DCAAE}{13} & \rqTSCell{DEF2F4}{1} & \rqTSZero & \rqTSZero & \rqTSTotal{14} & & \rqTSZero & \rqTSZero & \rqTSZero & \rqTSZero & \rqTSZero & \rqTSTotal{0} & & \rqTSZero & \rqTSZero & \rqTSZero & \rqTSZero & \rqTSZero & \rqTSTotal{0} \\
SF   & \rqTSCell{9BD9CB}{10} & \rqTSZero & \rqTSZero & \rqTSZero & \rqTSZero & \rqTSTotal{10} & & \rqTSCell{B2E1D8}{7} & \rqTSZero & \rqTSZero & \rqTSCell{DEF2F4}{1} & \rqTSZero & \rqTSTotal{8} & & \rqTSZero & \rqTSZero & \rqTSZero & \rqTSCell{DEF2F4}{1} & \rqTSZero & \rqTSTotal{1} & & \rqTSCell{DEF2F4}{1} & \rqTSZero & \rqTSZero & \rqTSZero & \rqTSZero & \rqTSTotal{1} \\
EmpR & \rqTSZero & \rqTSZero & \rqTSZero & \rqTSZero & \rqTSZero & \rqTSTotal{0} & & \rqTSCell{D6EFF0}{2} & \rqTSZero & \rqTSCell{C8EAE6}{4} & \rqTSZero & \rqTSZero & \rqTSTotal{6} & & \rqTSZero & \rqTSCell{DEF2F4}{1} & \rqTSZero & \rqTSZero & \rqTSZero & \rqTSTotal{1} & & \rqTSCell{DEF2F4}{1} & \rqTSCell{CFEDEB}{3} & \rqTSCell{DEF2F4}{1} & \rqTSZero & \rqTSZero & \rqTSTotal{5} \\
CO   & \rqTSCell{CFEDEB}{3} & \rqTSZero & \rqTSZero & \rqTSZero & \rqTSCell{DEF2F4}{1} & \rqTSTotal{4} & & \rqTSCell{DEF2F4}{1} & \rqTSCell{D6EFF0}{2} & \rqTSZero & \rqTSZero & \rqTSZero & \rqTSTotal{3} & & \rqTSZero & \rqTSCell{D6EFF0}{2} & \rqTSZero & \rqTSZero & \rqTSZero & \rqTSTotal{2} & & \rqTSZero & \rqTSCell{D6EFF0}{2} & \rqTSZero & \rqTSZero & \rqTSZero & \rqTSTotal{2} \\
TH   & \rqTSZero & \rqTSCell{D6EFF0}{2} & \rqTSZero & \rqTSZero & \rqTSZero & \rqTSTotal{2} & & \rqTSZero & \rqTSZero & \rqTSCell{DEF2F4}{1} & \rqTSCell{DEF2F4}{1} & \rqTSCell{DEF2F4}{1} & \rqTSTotal{3} & & \rqTSZero & \rqTSCell{C8EAE6}{4} & \rqTSZero & \rqTSZero & \rqTSZero & \rqTSTotal{4} & & \rqTSZero & \rqTSZero & \rqTSCell{D6EFF0}{2} & \rqTSZero & \rqTSZero & \rqTSTotal{2} \\
\cmidrule(lr){1-7}\cmidrule(lr){9-14}\cmidrule(lr){16-21}\cmidrule(lr){23-28}
Total & \rqTSTotal{56} & \rqTSTotal{2} & \rqTSTotal{6} & \rqTSTotal{11} & \rqTSTotal{7} & \rqTSGrandTotal{82} & & \rqTSTotal{33} & \rqTSTotal{23} & \rqTSTotal{14} & \rqTSTotal{18} & \rqTSTotal{2} & \rqTSGrandTotal{90} & & \rqTSTotal{5} & \rqTSTotal{37} & \rqTSTotal{2} & \rqTSTotal{3} & \rqTSTotal{0} & \rqTSGrandTotal{47} & & \rqTSTotal{14} & \rqTSTotal{9} & \rqTSTotal{12} & \rqTSTotal{1} & \rqTSTotal{0} & \rqTSGrandTotal{36} \\
\bottomrule
\end{tabular}
\end{adjustbox}
\end{table*}

Summing the bottom \emph{Tot} row across the four panels of \cref{tab:trigger-symptom} gives the symptom totals, and we therefore introduce the symptoms from most to least frequent. We abbreviate symptom categories as follows: SE = silent errors, Cr = crash, EIO = error in the output, RL = retry loop, and H = hang. The rows in the table represent triggering behaviors, which we introduce in RQ2.

\textbf{Silent Errors (108):} The agent appears to run normally and produces a fluent answer, action trace, or workspace change, but the result is incorrect and no exception, warning, or retry loop alerts the user. Unlike silent errors in traditional software, this symptom in an agent is often difficult to judge from the final response alone: the evidence needed by the oracle may lie in the tool log, workspace diff, or long interaction trajectory.

\begin{figure}[t]
\centering
\includegraphics[width=0.88\linewidth]{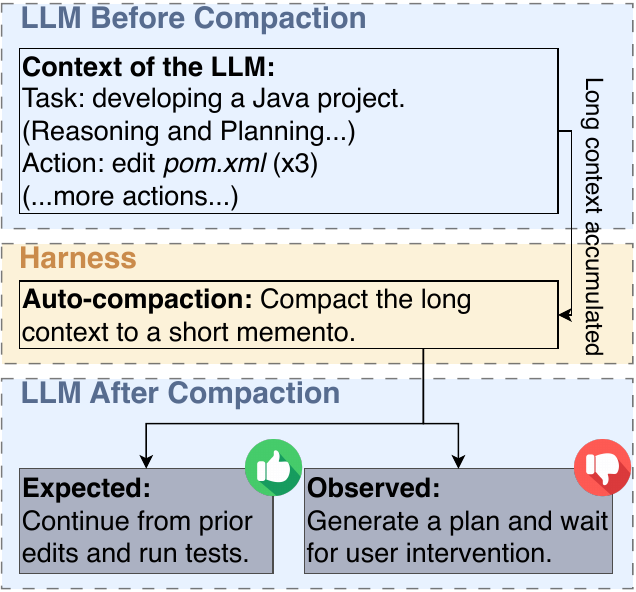}
\caption{Codex issue 5957: after auto-compaction, the agent should continue from prior edits but instead generates a new plan and waits for the user.}
\label{fig:codex5957}
\end{figure}

We highlight two representative cases of such issues. In Codex issue 5957~\cite{codex5957} (as shown in \cref{fig:codex5957}), the agent worked on a Java repository and was expected to continue editing files until the task was complete. After the agent had edited multiple files, the harness automatically compacted the long context into a short memento that omitted those edits. The expected behavior was to continue from the previous task state after compaction. Instead, the agent behaved as if it had just started: it forgot the previous edits, generated a fresh plan, and waited for user intervention. The symptom is silent because no crash or warning tells the user that the task state was lost; the user must inspect the chat history and workspace state to distinguish this premature ending from normal task completion.

Another example is Codex issue 6562~\cite{codex6562}, in which the agent claimed that it performed a sequence of verification steps: running \texttt{pnpm dev}, opening DevTools, clearing the network log, and observing the request payload. However, the user never granted it permission to access the relevant tools, and the harness had no record of such tool calls. The LLM's response is fluent and structured as a successful audit trail, so neither the harness nor the user receives any signal that the work never happened. To notice this failure, the user cannot rely on the final response alone; they must compare the agent's claimed actions with the configuration and tool-call log. 

The SE column of \cref{tab:trigger-symptom} contains 108 issues (56 in Codex, 33 in Gemini-CLI, 14 in CrewAI, and 5 in LangChain), making silent errors the largest symptom in our corpus. They are hard to detect because the agent's reply is fluent and self-consistent, leaving no exception or warning; the user only notices the discrepancy if they manually inspect the working trajectory (as in Codex 5957, illustrated in \cref{fig:codex5957}) or compare the agent's narrated steps against the record of actual tool calls (as in Codex 6562). These cases can also be hard to reproduce because reproduction may rely on long context, complex workspace state, and a particular sequence of LLM behaviors.

\textbf{Crash (71):} Exceptions are raised, and the whole agent session terminates or enters an unrecoverable error state, and users cannot continue working with the session. We illustrate the crash symptom with two representative cases that span the situation faced by both agent SDK users and agent product CLI users.

In LangChain issue 1358~\cite{langchain1358}, the user initializes a \texttt{ConversationChatAgent} with HuggingFace's \texttt{flan-t5-xl}, then the framework raises \texttt{ValueError: Could not parse LLM output: ``Assistant, how can I help you today?''}. 
Another example is Gemini-CLI issue 13292~\cite{gemini13292}, where a long agent session crashes mid-response with a backend 400 stating \texttt{``Please ensure that the number of function response parts is equal to the number of function call parts of the function call turn.''}. Once the error fires, every subsequent prompt returns the same wall of 400 messages, and the user reports there is no way to continue the conversation. Although the agent program does not terminate, we still consider this issue a crash because the session cannot continue, producing an effect similar to a crash in traditional software. 

Compared with agent products (Gemini-CLI and Codex), customized agents built with LangChain and CrewAI suffer from the Crash symptom more frequently in proportion (as shown in \cref{tab:trigger-symptom}). Crash accounts for 79\% of LangChain issues (37/47) and 25\% of CrewAI issues (9/36), versus 26\% of Gemini-CLI issues (23/90) and only 2\% of Codex issues (2/82).
Crash occurs in both traditional software projects and agents and is usually easy to detect. Some crash issues also attracted high engagement of users. LangChain issue 1358 accumulates 124 thumbs-up reactions and 81 comments, which is the highest-engagement issue in our corpus. Gemini-CLI 13292 is also the canonical issue for many duplicate reports, indicating that a crash is a significant symptom in agents.

\textbf{Error in the Output (34):} The session is still working, but it displays an error message instead of a normal AI message or tool message. Unlike silent errors, this symptom is directly noticeable because the error message can be distinguished from normal output.

\begin{figure}[t]
\centering
\includegraphics[width=\linewidth]{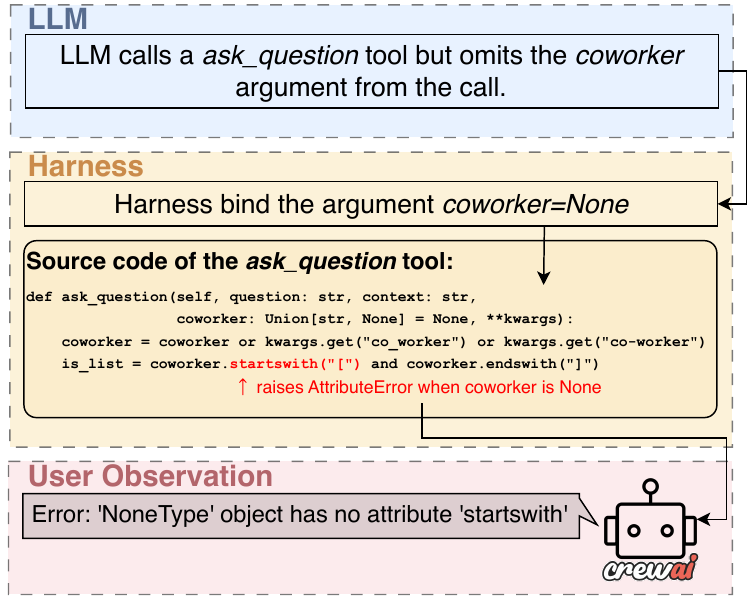}
\caption{CrewAI issue 668: an omitted tool argument becomes an error in the output.}
\label{fig:crewai668}
\end{figure}

For example, in CrewAI issue 668~\cite{crewai668} (\cref{fig:crewai668}), the LLM calls the \texttt{ask\_question} tool but omits the \texttt{coworker} argument. The harness binds this missing value as \texttt{None}; after the tool code calls \texttt{startswith} on it, the user only receives \texttt{'NoneType' object has no attribute 'startswith'}. Such a raw error message is not a useful signal for recovery and is confusing for users to debug.\looseness=-1

The EIO column of \cref{tab:trigger-symptom} contains 34 issues, concentrated in Gemini-CLI (14) and CrewAI (12), with fewer cases in Codex (6) and LangChain (2). EIO is easy to detect because the user sees a raw error message in place of normal output, but it is harder to reproduce because the occurrences are context-dependent and environment-dependent. We found in the comments of CrewAI issue 668 that the workarounds shared in the issue threads tend to help some users while failing for others, suggesting the uncertainty of this bug.

\textbf{Retry Loop (33):} The agent repeatedly takes the same or semantically equivalent action until it reaches an iteration/time limit or the user intervenes. Unlike a hang, the agent is still active but unproductive.

For example, in Gemini-CLI issue 5629~\cite{gemini5629}, the user asks Gemini to edit a file, and the agent invokes the \texttt{replace} tool with an \texttt{old\_string} that does not exactly match any substring on disk. The tool returns ``Failed to edit, 0 occurrences found for old\_string'', after which the agent re-reads the file, retries an edit with the same problem, fails again, and remains in this read-fail-retry cycle until the iteration limit is reached or the user closes the session. 

This differs from infinite loops in traditional software systems. On one hand, retry loops in agents waste user tokens without producing useful results, which makes users less tolerant of this symptom. On the other hand, the retry loop pollutes the agent's context, thus affecting subsequent work.

The RL column of \cref{tab:trigger-symptom} contains 33 issues, concentrated in Gemini-CLI (18) and Codex (11), with only 3 in LangChain and 1 in CrewAI. This distribution shows that retry loops are much more common in the two CLI products than in the two agent frameworks in our corpus.
Detection of RL is straightforward when the LLM repeats the same action verbatim, but can be challenging when it repeats semantically similar actions. 

\textbf{Hang (9):} The agent becomes unresponsive or remains in a working/waiting state without visible progress. Unlike a retry loop, a hang does not repeatedly execute the same actions and may produce no action at all.

The H column of \cref{tab:trigger-symptom} contains only 9 issues, all from the two CLI products: 7 in Codex and 2 in Gemini-CLI. However, it is still worth separating from other symptoms because the user's session is blocked without visible progress. For example, in Codex issue 4337~\cite{codex4337} (\cref{fig:codex4337}), the agent stays in the working state after a command timeout instead of moving to a failure state and trying other commands. As \cref{fig:codex4337} shows, after timeout, the wrapper kills the main process, but a surviving child process keeps running and hold stdout/stderr pipes open. However, the harness still waits in the working state. We explain the cause of this hang in RQ2.

\noindent\textbf{RQ1 Finding:} Agent failures are not limited to explicit crashes or raw errors. The most common symptom is silent errors, where users must inspect external evidence such as tool logs, workspace changes, or long trajectories to notice the failure. Crash and EIO are easier to observe but still leave localization difficult, while retry loops and hangs waste tokens, time, or an entire session. These symptoms motivate RQ2, where we analyze the LLM behaviors that trigger them.

\subsection{RQ2: Triggering Behaviors Noticed by Users}
\label{sec:rq2}
RQ2 asks which LLM behaviors expose the user-visible symptoms described in RQ1. The behavior itself is not necessarily faulty; an AR bug manifests when the harness parses, executes, or compacts that behavior in a way that produces a symptom. Thus, this subsection focuses on the interaction between model decisions and harness control flow, rather than assigning the fault solely to either the backend LLM or the harness.

We introduce these behaviors from more to less frequent in \cref{tab:trigger-symptom}, which abbreviates them as follows: INC = task instruction non-compliance, UTA = unexpected tool arguments, MTC = message template conflict, IHI = immature harness interface, SF = statement fabrication, EmpR = empty response, CO = context overflow, and TH = tool hallucination. For each behavior, we describe the triggering behavior noticed by users, use a representative issue to show how the harness turns it into a symptom, and summarize its distribution across symptoms.

\textbf{Task Instruction Non-compliance (67):} We label a behavior as task instruction non-compliance when the model ignores task instructions or system prompts while still producing syntactically acceptable output. If the output conflicts with a required template, we label it as message template conflict (MTC), which we discuss below. 
For example, in Codex issue 13491~\cite{codex13491} (\cref{fig:codex13491}), an orchestrator calls the spawn-agent tool with \texttt{fork\_context=true}. As \cref{fig:codex13491} shows, the harness builds the sub-agent context by forking the orchestrator context and appending a handoff message that tells the sub-agent to treat the next user message as its new task. However, the sub-agent treats the inherited context as an instruction for itself, ignores the appended handoff, and attempts to spawn sub-agents recursively.
Instruction non-compliance usually manifests as silent errors, as the INC row of \cref{tab:trigger-symptom} shows that 54 of 67 issues (81\%) surface as SE. This is because when a model ignores an instruction, it can still return fluent and well-formed text that satisfies the harness's surface contract. The remaining non-SE cases arise only when the ignored instruction changes control flow, such as causing repeated delegation in Codex issue 13491~\cite{codex13491}.

\textbf{Unexpected Tool Arguments (62):} Invoking tools to complete tasks is a crucial feature of agents, and the tool execution starts from the tool arguments provided from LLMs.
UTA occurs when the model calls valid tools with existing arguments, but the argument \emph{value} is in unexpected format or datatype, or the agent framework fail to handle the argument. If the model uses a wrong tool name or argument name, we label it as TH instead. 
For example, in CrewAI issue 668~\cite{crewai668} (\cref{fig:crewai668}), the LLM invokes the \texttt{ask\_question} tool but omits the required \texttt{coworker} value. The figure shows the harness binds the missing value as \texttt{None}, dispatches it to the tool body, and the tool dereferences it through \texttt{coworker.startswith(...)}, producing a raw Python \texttt{AttributeError}.
Another example is Gemini-CLI issue 3037~\cite{gemini3037}, where the argument value produced by the LLM is a commit message containing backticks. When the harness passes it to \texttt{git commit} through the shell, the shell interprets the backticks as command substitution and drops the wrapped text. In this case, the argument value is a valid string, but the failure comes from how the harness handles it.

Unlike INC, UTA has no dominant symptom: crash (19), retry loop (19), and EIO (14) each account for a substantial fraction of the 62 cases. The split tracks the harness's tool-dispatch design. In \cref{tab:trigger-symptom}, the LangChain panel's UTA row has 11 cases in the Cr column out of 12 UTA cases, because LangChain's wrappers often propagate the underlying Python exception. In the Gemini-CLI panel, the UTA row has 13 cases in the RL column out of 29 UTA cases, because Gemini-CLI often returns the tool error to the model as an observation and allows repair attempts. In the CrewAI panel, the UTA row has 5 cases in the EIO column out of 7 UTA cases, because CrewAI often surfaces the raw Python error as the tool output. The same class of LLM behaviors therefore manifests as different symptoms depending on the layer of the harness that catches the unexpected arguments. 

\textbf{Message Template Conflict (49):} MTC occurs when the model's output does not match the message or action template that the harness parser expects. This differs from INC because the key failure is not simply disobeying a task instruction, but crossing a parser boundary. For example, in LangChain issue 1358~\cite{langchain1358}, a \texttt{ConversationChatAgent} using HuggingFace's \texttt{flan-t5-xl} responds with conversational text instead of the ReAct-style markers that the parser expects, such as \texttt{Thought:}, \texttt{Action:}, and \texttt{Action Input:}. The parser cannot extract the next action and raises \texttt{ValueError: Could not parse LLM output}.

The case illustrates how a parser contract designed around a particular action format can fail when users plug in a model that emits a different conversational format. MTC maps to symptoms according to the strictness of parser and recovery policy in the harness. Across the corpus, it manifests as crash (21), silent errors (19), EIO (7), and retry loop (2). This behavior is concentrated in LangChain and CrewAI, where users combine framework parsers with diverse backend models to build their customized agents, which may not have parsers and recovery policies that are robust to formats of all LLMs. In LangChain, 18 of 23 MTC cases manifest as crash because parser failures are raised as exceptions. In CrewAI, the symptoms triggered by MTC spread across SE, crash, EIO, and RL, indicating that different layers can catch or expose the mismatch differently.
For agent CLIs, Codex and Gemini-CLI show fewer MTC cases in our corpus. One plausible reason is that their harnesses and default model families (i.e., GPT and Gemini series) are co-designed around similar interaction formats.

\begin{figure}[t]
\centering
\includegraphics[width=\linewidth]{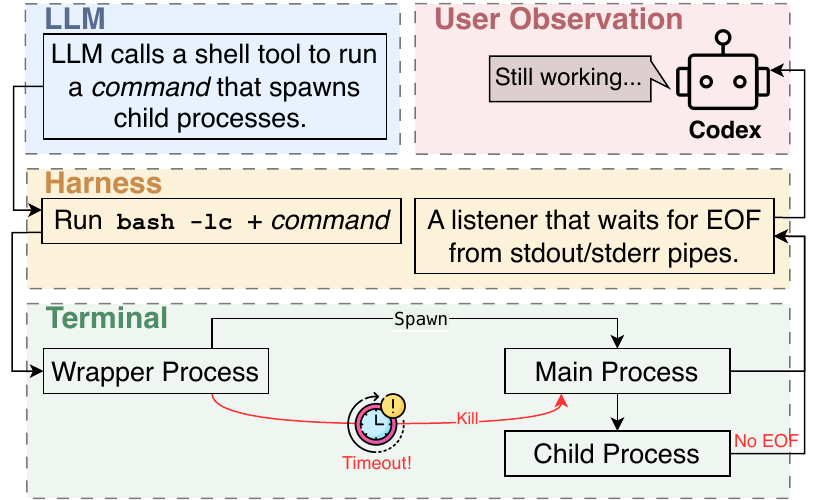}
\caption{Codex issue 4337: after timeout, the wrapper kills the main process, but a child process keeps running and leaves Codex still working.}
\label{fig:codex4337}
\end{figure}

\textbf{Immature Harness Interface (23):} IHI captures cases where the LLM takes a valid action, but the harness interface cannot coordinate, terminate, or recover from that action correctly. These cases often involve gaps between what the model attempts to do and what the surrounding agent runtime can safely support. For example, in Codex issue 4337~\cite{codex4337} (\cref{fig:codex4337}), the LLM invokes a shell tool, and the harness runs the command through its standard \texttt{bash -lc} wrapper. The command spawns a sub-process and then times out. The wrapper kills the main process, but leaves a child process alive and keeps stdout or stderr pipes open. The harness therefore waits for an EOF that never arrives, and the agent is stuck in the ``working'' state. Here the triggering action is an ordinary shell invocation, but the symptom comes from the process-management interface in the harness. 

\textbf{Statement Fabrication (20):} SF occurs when the model claims an observation or state that is inconsistent with facts on the user's device. For exmaple, in Codex issue 6562~\cite{codex6562}, the agent fabricated that it performed verification steps using DevTools, but the user had never granted access to the relevant tools and the harness had no corresponding tool-call record. Another exmaple is CrewAI issue 3154~\cite{crewai3154}, an agent configured with a \texttt{WebSearchTool} returned content that looks like web search results, yet the \texttt{WebSearchTool.run()} was never invoked. Actually, the LLM fabricates the web search results directly instead of calling a real tool. 
These claims are difficult to distinguish because users must compare the model's narration with configurations, tool logs, network activity, or workspace state.

SF overwhelmingly leads to silent errors: 18 of 20 cases (90\%) manifest as SE. The reason is that the fabricated statement is itself a fluent answer or progress report, so no exception is produced. The two non-SE cases become retry loops when the fabricated statement triggers subsequent repeated actions.

\textbf{Empty Response (12):} EmpR denotes literally empty model responses, such as empty content or empty decision, not merely low-quality or unhelpful content. In Gemini-CLI issue 7223~\cite{gemini7223}, the backend model returns an empty completion, often after a \texttt{ReadManyFiles} call that loads a large set of files into context. Before a harness-side workaround, the Gemini-CLI silently drops the user's most recent prompt and continues as if it had received no model content. After the workaround, the harness automatically retries the request. If all retries fail, it responds the error \texttt{''The model failed to respond after multiple attempts.''} to the user.

The symptom of EmpR depends on how the harness handles the empty payload. Across the corpus, EmpR manifests as EIO (5), crash (4), and silent errors (3). CrewAI and LangChain usually treat an empty response as an unrecoverable value error, and produce the crash symptom. Gemini-CLI more often shows the corresponding error message to the user.

\textbf{Context Overflow (11):} CO covers failures triggered by long memory content. Accumulated long messages generated by LLMs or overly large tool results can lead to a long memory content. CO manifests mainly as crash (6) or silent errors (4). 
For example, Codex issue 5957~\cite{codex5957} (\cref{fig:codex5957}) mentioned in RQ1 falls into silent errors. In this issue, long context triggers automatic compaction, but the compacted memento omits prior file edits, so the agent generates a new plan instead of continuing from the prior edits. 
However, in CrewAI issue 3843~\cite{crewai3843}, the user asks for a single Elasticsearch record, but the agent retrieves ten thousand records; once the tool output is appended to context, the next model request exceeds the token window and the agent crashes.

This is a case where the same LLM behavior can become either an explicit failure or a plausible but wrong output, indicating that the behavior-symptom mapping is useful not only for root-cause analysis, but also for designing robust harness.

\textbf{Tool Hallucination (11):} TH occurs when the model use a tool or tool argument that the tool registry does not define, or refers to an existing tool or argument under an incorrect name. This category concerns names, not argument values.

The 11 TH cases mainly manifest as crash (6) and EIO (3). The same incorrect name can crash when the harness treats it as an unresolved registry lookup (in LangChain and Codex), or appear as EIO when the harness returns an error message as output (in Gemini-CLI and CrewAI).

\noindent\textbf{RQ2 Finding:}
The most common triggering behaviors are INC (67), UTA (62), and MTC (49). They expose two different failure paths.
INC and SF often preserve the harness's surface contract, so they mostly surface as silent errors.
Other triggering behaviors, in contrast, usually violate interaction rules defined in the harness. Their symptoms therefore depend on how the harness validates, propagates, retries, or recovers from the behavior.
Thus, an AR bug cannot be understood from the LLM behavior alone: the same behavior can surface as different symptoms under different harness handling.
This interaction between LLM and harness makes repair-target attribution difficult, because the triggering behavior is produced by the LLM while the observed symptom is shaped by the harness.

\subsection{RQ3: Attitudes from Users and Developers Toward Bug Repair}
\label{sec:rq3}

To answer RQ3, we investigate which component each issue discussion targets for repair: the agent harness, the backend LLM, or both. We infer this target from repository activity, including pull requests (PRs) and comments by users and project developers.

We categorize these actions into three groups: harness-side fixes, model-side fixes, and ambiguous or no engagement. Each issue receives one developer-side code and one user-side code; when multiple signals appear, we apply the rules in the order described below.
For developers, an issue with a linked non-bot PR merged into the project is marked as \texttt{D-HarnessPR} (abbreviated as \texttt{H-PR}), because a merged PR indicates that the developers modified the harness.
Otherwise, if a developer explicitly mentioned in the issue comments that the issue will be handled by improving the backend LLM, the issue is marked as \texttt{D-LLM} (abbreviated as \texttt{LLM}). 
An issue with developer participation yet without a clear fix location is marked as \texttt{D-Ambiguous} (abbreviated as \texttt{Ambig.}); an issue without human developer participation is marked as \texttt{D-NoEngagement} (abbreviated as \texttt{No Eng.}). 
We similarly extract user actions or suggestions from user-created PRs, issue bodies, and comments. We mark issues with user-created PRs as \texttt{U-HarnessPR} (abbreviated as \texttt{H-PR}).
If there is no user-created PR, we mark the user complaints that point to the backend LLM, the harness, or both layers as \texttt{U-ComplainLLM} (abbreviated as \texttt{C-LLM}), \texttt{U-ComplainHarness} (abbreviated as \texttt{C-Harness}), or \texttt{U-ComplainBoth} (abbreviated as \texttt{C-Both}).
Finally, issue reports that only describe observations without expressing a clear position are marked as \texttt{U-NoPosition} (abbreviated as \texttt{No Pos.}). As in RQ1 and RQ2, two human authors independently annotated the 255 issues following this guidance and resolved disagreements through discussion. Bots are not treated as developers or users, and PRs or comments created or merged only by bots are excluded from these rules.

\begin{table}[t]
\centering
\caption{Developer action by triggering behavior. H-PR = merged harness PR; LLM = developer attributes the issue to the backend LLM; Ambig. = developer engages without a clear repair target; No Eng. = no developer engagement.}
\small
\label{tab:rq3-dev-by-trigger}
\begin{adjustbox}{max width=\textwidth}
\begin{tabular}{lrrrrr}
\toprule
Trigger & H-PR & LLM & Ambig. & \makecell{No\\Eng.} & Total \\
\midrule
INC & 11 & 22 & 24 & 10 & 67 \\
UTA & 22 & 5 & 28 & 7 & 62 \\
MTC & 18 & 5 & 17 & 9 & 49 \\
IHI & 2 & 0 & 18 & 3 & 23 \\
SF & 1 & 5 & 11 & 3 & 20 \\
EmpR & 1 & 1 & 10 & 0 & 12 \\
CO & 2 & 1 & 3 & 5 & 11 \\
TH & 5 & 0 & 5 & 1 & 11 \\
\midrule
Total & 62 & 39 & 116 & 38 & 255 \\
\bottomrule
\end{tabular}
\end{adjustbox}
\end{table}

\begin{figure}[t]
\centering
\definecolor{rq3UserPR}{RGB}{31,119,180}
\definecolor{rq3UserBoth}{RGB}{148,103,189}
\definecolor{rq3UserH}{RGB}{44,160,144}
\definecolor{rq3UserL}{RGB}{231,138,60}
\definecolor{rq3UserNone}{RGB}{140,140,140}
\definecolor{rq3DevAgree}{RGB}{44,160,70}
\definecolor{rq3DevLLM}{RGB}{227,119,60}
\definecolor{rq3DevAmb}{RGB}{210,180,70}
\definecolor{rq3DevNone}{RGB}{140,140,140}
\begin{tikzpicture}[font=\footnotesize, x=0.58cm, y=0.44cm]
\fill[rq3UserPR] (0.000,0.798) rectangle (0.800,4.500);
\node[anchor=east, font=\footnotesize] at (-0.120,2.649) {H-PR (114)};
\fill[rq3UserBoth] (0.000,0.456) rectangle (0.800,0.618);
\node[anchor=east, font=\footnotesize] at (-0.120,0.537) {C-Both (5)};
\fill[rq3UserH] (0.000,-1.315) rectangle (0.800,0.276);
\node[anchor=east, font=\footnotesize] at (-0.120,-0.520) {C-Harness (49)};
\fill[rq3UserL] (0.000,-1.722) rectangle (0.800,-1.495);
\node[anchor=east, font=\footnotesize] at (-0.120,-1.609) {C-LLM (7)};
\fill[rq3UserNone] (0.000,-4.500) rectangle (0.800,-1.902);
\node[anchor=east, font=\footnotesize] at (-0.120,-3.201) {No Pos. (80)};
\fill[rq3DevAgree] (8.000,2.443) rectangle (8.800,4.500);
\node[anchor=west, font=\footnotesize] at (8.920,3.472) {H-PR (62)};
\fill[rq3DevLLM] (8.000,0.969) rectangle (8.800,2.263);
\node[anchor=west, font=\footnotesize] at (8.920,1.616) {LLM (39)};
\fill[rq3DevAmb] (8.000,-3.059) rectangle (8.800,0.789);
\node[anchor=west, font=\footnotesize] at (8.920,-1.135) {Ambig. (116)};
\fill[rq3DevNone] (8.000,-4.500) rectangle (8.800,-3.239);
\node[anchor=west, font=\footnotesize] at (8.920,-3.870) {No Eng. (38)};
\fill[rq3UserPR, opacity=0.45] (0.800,4.500) .. controls (4.040,4.500) and (4.760,4.500) .. (8.000,4.500) -- (8.000,3.073) .. controls (4.760,3.073) and (4.040,3.104) .. (0.800,3.104) -- cycle;
\fill[rq3UserPR, opacity=0.45] (0.800,3.104) .. controls (4.040,3.104) and (4.760,2.263) .. (8.000,2.263) -- (8.000,2.130) .. controls (4.760,2.130) and (4.040,2.974) .. (0.800,2.974) -- cycle;
\fill[rq3UserPR, opacity=0.45] (0.800,2.974) .. controls (4.040,2.974) and (4.760,0.789) .. (8.000,0.789) -- (8.000,-1.002) .. controls (4.760,-1.002) and (4.040,1.220) .. (0.800,1.220) -- cycle;
\fill[rq3UserPR, opacity=0.45] (0.800,1.220) .. controls (4.040,1.220) and (4.760,-3.239) .. (8.000,-3.239) -- (8.000,-3.671) .. controls (4.760,-3.671) and (4.040,0.798) .. (0.800,0.798) -- cycle;
\fill[rq3UserBoth, opacity=0.45] (0.800,0.618) .. controls (4.040,0.618) and (4.760,2.130) .. (8.000,2.130) -- (8.000,2.064) .. controls (4.760,2.064) and (4.040,0.553) .. (0.800,0.553) -- cycle;
\fill[rq3UserBoth, opacity=0.45] (0.800,0.553) .. controls (4.040,0.553) and (4.760,-1.002) .. (8.000,-1.002) -- (8.000,-1.069) .. controls (4.760,-1.069) and (4.040,0.488) .. (0.800,0.488) -- cycle;
\fill[rq3UserBoth, opacity=0.45] (0.800,0.488) .. controls (4.040,0.488) and (4.760,-3.671) .. (8.000,-3.671) -- (8.000,-3.704) .. controls (4.760,-3.704) and (4.040,0.456) .. (0.800,0.456) -- cycle;
\fill[rq3UserH, opacity=0.45] (0.800,0.276) .. controls (4.040,0.276) and (4.760,3.073) .. (8.000,3.073) -- (8.000,2.874) .. controls (4.760,2.874) and (4.040,0.081) .. (0.800,0.081) -- cycle;
\fill[rq3UserH, opacity=0.45] (0.800,0.081) .. controls (4.040,0.081) and (4.760,2.064) .. (8.000,2.064) -- (8.000,1.633) .. controls (4.760,1.633) and (4.040,-0.341) .. (0.800,-0.341) -- cycle;
\fill[rq3UserH, opacity=0.45] (0.800,-0.341) .. controls (4.040,-0.341) and (4.760,-1.069) .. (8.000,-1.069) -- (8.000,-1.865) .. controls (4.760,-1.865) and (4.040,-1.120) .. (0.800,-1.120) -- cycle;
\fill[rq3UserH, opacity=0.45] (0.800,-1.120) .. controls (4.040,-1.120) and (4.760,-3.704) .. (8.000,-3.704) -- (8.000,-3.903) .. controls (4.760,-3.903) and (4.040,-1.315) .. (0.800,-1.315) -- cycle;
\fill[rq3UserL, opacity=0.45] (0.800,-1.495) .. controls (4.040,-1.495) and (4.760,2.874) .. (8.000,2.874) -- (8.000,2.841) .. controls (4.760,2.841) and (4.040,-1.528) .. (0.800,-1.528) -- cycle;
\fill[rq3UserL, opacity=0.45] (0.800,-1.528) .. controls (4.040,-1.528) and (4.760,1.633) .. (8.000,1.633) -- (8.000,1.467) .. controls (4.760,1.467) and (4.040,-1.690) .. (0.800,-1.690) -- cycle;
\fill[rq3UserL, opacity=0.45] (0.800,-1.690) .. controls (4.040,-1.690) and (4.760,-3.903) .. (8.000,-3.903) -- (8.000,-3.936) .. controls (4.760,-3.936) and (4.040,-1.722) .. (0.800,-1.722) -- cycle;
\fill[rq3UserNone, opacity=0.45] (0.800,-1.902) .. controls (4.040,-1.902) and (4.760,2.841) .. (8.000,2.841) -- (8.000,2.443) .. controls (4.760,2.443) and (4.040,-2.292) .. (0.800,-2.292) -- cycle;
\fill[rq3UserNone, opacity=0.45] (0.800,-2.292) .. controls (4.040,-2.292) and (4.760,1.467) .. (8.000,1.467) -- (8.000,0.969) .. controls (4.760,0.969) and (4.040,-2.779) .. (0.800,-2.779) -- cycle;
\fill[rq3UserNone, opacity=0.45] (0.800,-2.779) .. controls (4.040,-2.779) and (4.760,-1.865) .. (8.000,-1.865) -- (8.000,-3.059) .. controls (4.760,-3.059) and (4.040,-3.948) .. (0.800,-3.948) -- cycle;
\fill[rq3UserNone, opacity=0.45] (0.800,-3.948) .. controls (4.040,-3.948) and (4.760,-3.936) .. (8.000,-3.936) -- (8.000,-4.500) .. controls (4.760,-4.500) and (4.040,-4.500) .. (0.800,-4.500) -- cycle;
\node[anchor=south, font=\bfseries\footnotesize] at (0.400,4.900) {User};
\node[anchor=south, font=\bfseries\footnotesize] at (8.400,4.900) {Developer};
\end{tikzpicture}
\caption{Repair target codes by user and developer. Ribbon width indicates issue count. H-PR = harness PR; C-Both = both layers; C-Harness = harness; C-LLM = LLM; No Pos. = no user position; Ambig. = ambiguous developer position; No Eng. = no developer engagement.}
\label{fig:rq3-sankey}
\end{figure}
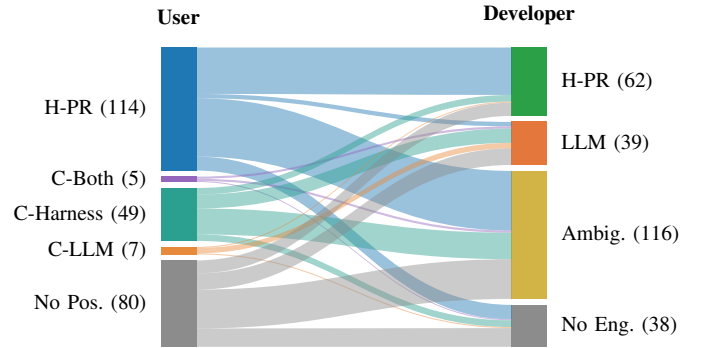

The Sankey diagram in \Cref{fig:rq3-sankey} connects each issue's user-side actions with its developer-side actions. Each ribbon shows whether a user-side position leads to a consistent or mismatched developer-side position, and whether user-engaged issues get responses from developers. 
We organize the result in this way to facilitate RQ3 to analyze each issue from a two-sided perspective, not only the distributions.

\textbf{User and developer positions.}
\Cref{fig:rq3-sankey} shows that, on the user axis, 114 of 255 issues (45\%) have user-created PRs for the agent harness (\texttt{U-HarnessPR}), 49 (19\%) complain about the harness (\texttt{U-ComplainHarness}), 7 (3\%) complain about the backend LLM (\texttt{U-ComplainLLM}), and 5 (2\%) complain about both layers (\texttt{U-ComplainBoth}); the remaining 80 issues (31\%) describe only the symptom without taking a position (\texttt{U-NoPosition}). 
On the developer axis, 62 issues (24\%) end with a merged harness PR (\texttt{D-HarnessPR}), 39 (15\%) have a developer explicitly attributing the failure to the model (\texttt{D-LLM}), 116 (46\%) have developers engaged by requesting more information or closing a PR, instead of committing to a fix location (\texttt{D-Ambiguous}), and 38 (15\%) have no developer engagement (\texttt{D-NoEngagement}). 
Overall, in our corpus, user-side actions or complaints most often target the harness: 168 of 255 issues (66\%) include a user-created PR or a harness-side complaint, while the largest developer-side group is ambiguous (116 of 255 issues, 46\%).

Considering that user-created PRs clearly indicate a harness-side action, we further inspect the user-side H-PR branch in \cref{fig:rq3-sankey}. 
Among the 114 issues in which users contributed PRs, only 43 (38\%) have been merged, and another 4 (4\%) drew explicit attribution to the LLM by developers. Only these 47 issues (41\%) reached a clear repair outcome under our rubric. The remaining issues stall in different ways: developers may hold a PR open pending more information or further improvements, close stale PRs, or leave the issue managed by Bots. 
As discussed in RQ1, reproducing agent bugs and designing test oracles can be difficult, which may increase the pressure on developers to handle PRs. This partially explains why a harness-side contribution does not always lead to an accepted repair.

\textbf{Mismatches between user and developer.}
\begin{table}[t]
\centering
\caption{User--developer mismatched issues by triggering behavior.}
\label{tab:rq3-mismatch-by-trigger}
\small
\begin{adjustbox}{max width=\columnwidth}
\begin{tabular}{lrrr}
\toprule
Trigger & Mismatch & Total & Rate \\
\midrule
INC  & 10 & 67 & 15\% \\
UTA  &  3 & 62 &  5\% \\
MTC  &  3 & 49 &  6\% \\
IHI  &  0 & 23 &  0\% \\
SF   &  2 & 20 & 10\% \\
EmpR &  1 & 12 &  8\% \\
CO   &  0 & 11 &  0\% \\
TH   &  0 & 11 &  0\% \\
\midrule
\textbf{Total} & \textbf{19} & \textbf{255} & \textbf{7\%} \\
\bottomrule
\end{tabular}
\end{adjustbox}
\end{table}

\Cref{fig:rq3-sankey} indicates a clear mismatch between users' suggestions and developers' repair targets, i.e., a non-trivial number of issues flow from \texttt{U-HarnessPR} or \texttt{U-ComplainHarness} to \texttt{D-LLM}. It suggests a phenomenon that even when users have suggested harness-side repair solutions, developers sometimes still explicitly state that they will resolve the issue by improving the LLM.
\Cref{tab:rq3-mismatch-by-trigger} further shows that the mismatch concentrates on \emph{Task Instruction Non-compliance} (INC) and \emph{Statement Fabrication} (SF), which have a mismatched issue rate greater than 10\%. Therefore, we further conduct a case study to understand the users' and developers' attitudes in both categories.

For INC, Codex~5807~\cite{codex5807} reports that Codex overwrites the user's changes to a file without requesting permission. The user proposed harness-side suggestions, such as recording and checking the version of files before applying patches from the LLM. However, the OpenAI developer responded that ``We've tried to do what you've proposed... This needs to be addressed at the model level." 
Although the user proposed more harness-side workarounds afterward, the developer has not responded as of the time of this paper's completion.

SF also leads to a high mismatching rate. Codex~10828~\cite{codex10828} is a representative issue, during which the agent claimed it was still working on the task, for example, ``I'm continuing now..., and will return results only once complete". But the turn had silently ended when the LLM was generating the response, and subsequent work was never performed. 
A user proposed a harness-side fix, asking the CLI to notify the user whenever a run stops unexpectedly, while the developer left a comment, stating that there is not much they can do except improve the backend LLM via training.

\textbf{Merged PRs reveal harness-side mitigation patterns.}
Beyond how developers respond, we examine how merged harness PRs mitigate these behaviors in our corpus. As shown in \cref{tab:rq3-dev-by-trigger}, merged harness PRs are concentrated in \emph{Unexpected Tool Arguments} (22), \emph{Message Template Conflict} (18), \emph{Task Instruction Non-compliance} (11), and \emph{Tool Hallucination} (5). We also conduct a further case study on these four behaviors to summarize the representative mitigation patterns.

\textbf{\emph{Unexpected Tool Arguments}}: The UTA row in \cref{tab:rq3-dev-by-trigger} contains the largest number of merged harness PRs, and these fixes commonly validate the model's tool calls before execution. For example, the merged fix for the \texttt{read\_file} crash in Gemini-CLI~533~\cite{gemini533} rejects an undefined path. However, when tool contracts are specified through natural-language descriptions or built-in types in the host programming language, it remains challenging to validate the arguments provided by LLMs before execution. 

\textbf{\emph{Message Template Conflict}}: The MTC row in \cref{tab:rq3-dev-by-trigger} also shows many merged harness PRs, often through more tolerant parsing or prompt steering. For example, LangChain~12077~\cite{langchain12077} strips the spurious \texttt{SQLQuery:} prefix the model prepends before the query runs, and LangChain~5163~\cite{langchain5163} appends a reminder that steers the router to wrap its output in a JSON code block. 

\textbf{\emph{Task Instruction Non-compliance}}: The INC row in \cref{tab:rq3-dev-by-trigger} includes 11 merged harness PRs but also 22 explicit model-side attributions. Its merged harness fixes span three mechanisms. When the model disobeys an instruction related to the user's customized output format, tolerant parsing mitigates it. For example, in LangChain~11408~\cite{langchain11408}, where the model was asked for a \texttt{YES} or \texttt{NO} answer but replied with words with similar meanings, the harness loosens the boolean parser to accept ``Not relevant (NO)''. When the model violates a mode or policy, the PRs strengthen the prompt. When INC behavior shows up as looping or repeated work, the PRs add runtime guards, such as the infinite-loop protection merged in Gemini-CLI~1484~\cite{gemini1484}. For other INC behaviors, the fix is still contested between users and developers, as discussed above.

\textbf{\emph{Tool Hallucination}}: For TH, \cref{tab:rq3-dev-by-trigger} shows 5 merged harness PRs among 11 issues, and the direct mitigation is to check tool names against the registry. LangChain~34910~\cite{langchain34910} raises a clear exception when the model calls a tool that does not exist. 

As the mismatch cases above show, the fix for \emph{statement fabrication} is also contested. There are too few merged PRs for the rest behaviors to summarize a representative fixing method. 

\textbf{RQ3 Finding.} Among our collected issue reports, users prefer taking harness-directed actions or complain about the harness, while developers often respond without a clear repair target or attribute some behaviors to the backend model. Accepted harness PRs mitigate AR bugs through validation, tolerant parsing, prompt steering, registry checks, and runtime guards. However, INC and SF still expose contested repair targets. The results of RQ3 suggest that maintaining agents would benefit from fault-localization support and reproducible tests that distinguish model limitations from harness robustness gaps.

\section{Discussion}
\label{sec:discussion}

\subsection{Practical Advice for Users and Developers}
Our taxonomy may serve as a checklist for the maintenance of agent products. When users report a failure, the symptom taxonomy may guide them to systematically describe what they observed: silent errors, a crash, an error in the output, a retry loop, or a hang. The triggering-behavior taxonomy then helps users and project developers collect the evidence needed for reproduction, such as the model output, tool arguments, context state, or tool-call log that exposed the failure. Finally, the repair-target taxonomy helps project developers decide whether the issue record supports a harness-side fix, a backend-model fix, or both. In this way, the taxonomy turns vague reports such as ``the agent failed'' into structured maintenance evidence: what failed, which LLM behavior exposed the failure, and where the repair is justified. Agent project maintainers can build their issue report template and bug-repairing workflow around these three RQs, and users can use the taxonomy to provide more actionable bug reports.

\subsection{Research Opportunities for Oracles and Reproduction}
The challenges of oracle design and reproduction observed in our study also suggest research opportunities. A test oracle for agents cannot solely rely on the final response, especially for silent errors and statement fabrication symptoms. Future work can design trace-based oracles that check whether an agent's claimed behavior or state is supported by real trajectory records. Reproducing AR bugs also needs new support. Existing work on structural testing of agents suggests using traces and mocks to make agent behavior reproducible and simplify the test environment~\cite{kohl2025structural}. AR bugs are suitable for this direction because their manifestations depend on specific LLM behaviors. Beyond replaying user inputs, tests should preserve or mock the triggering behavior of LLMs, including the specific tool arguments, template-breaking message, or long context that exposed the failure. Our triggering-behavior taxonomy can guide which LLM response or interaction state should be mocked, so developers can evaluate harness repairs without depending on the backend model to regenerate the bug-triggering response.

\section{Threats to Validity}
\label{sec:threats}

First, our study involves manual inspection of bug reports. These subjective steps may be biased because we manually infer symptoms, triggering LLM behaviors, and repair targets from issue reports, which can be incomplete and interpreted by annotators differently. To reduce this threat, two annotators worked independently and resolved conflicting cases through discussion until consensus was reached.

Second, it is unclear to what extent our findings can be generalized to other agents. We study 255 issues from four agent projects that are popular and publicly disclose issue trackers and source code of the harness.
However, they may not cover agents with different users or harness designs. Expanding the corpus is difficult because many popular agents, such as Claude Code, are closed-source or lack public trackers of issues and PRs. We cannot validate how users and developers detect, locate, and fix bugs without such information.

\section{Related Work}
\label{sec:relatedwork}

\subsection{LLM Agents and Harness Engineering}
Early LLM applications mainly used a model as a single-turn chat or completion engine. Prompting techniques improve this interaction by providing examples~\cite{brown2020fewshot}, teaching models to follow instructions~\cite{ouyang2022training}, or eliciting step-by-step reasoning~\cite{wei2022chain}. 
However, many tasks require coordination beyond one response. Multi-agent systems therefore assign different roles to LLMs. AutoGen supports conversable-agent programming~\cite{wu2023autogen}. MetaGPT encodes standard operating procedures~\cite{hong2024metagpt}. ChatDev organizes software-development roles through chat chains~\cite{qian2024chatdev}. These systems still need an execution substrate when agents must use tools, observe results, and continue working. Loop-based agents provide this substrate. ReAct interleaves reasoning and acting~\cite{yao2023react}. ToolLLM studies API use~\cite{qin2024toolllm}. SWE-agent builds an agent-computer interface for software engineering~\cite{yang2024sweagent}. In such agents, the harness renders prompts, parses actions, dispatches tools, manages context, and coordinates recovery~\cite{xi2023rise,yang2024sweagent,meng2026harness}. Our work studies the failure introduced by this model-harness architecture.

\subsection{Testing and Debugging of AI Systems}
AI-based systems create maintenance problems that differ from traditional software. ML pipelines accumulate hidden technical debt~\cite{sculley2015debt}. Production ML systems require new engineering practices~\cite{amershi2019se4ml}. Deep-learning frameworks also expose distinctive bug patterns~\cite{islam2019dlbugs,humbatova2020taxonomy}. Testing work responds with input generation~\cite{pei2017deepxplore}, cross-backend validation~\cite{pham2019cradle}, and behavioral tests~\cite{ribeiro2020checklist}. For LLMs, existing work evaluates instruction following~\cite{ouyang2022training}, factuality~\cite{lin2022truthfulqa,min2023factscore}, hallucination~\cite{manakul2023selfcheckgpt,li2023halueval}, and faithfulness~\cite{maynez2020faithfulness}. These studies motivate our focus on model behaviors and observable symptoms. Our study further asks where the repair should happen when such behavior is wrapped by an agent harness. This question also connects to bug-report research on report quality~\cite{bettenburg2008goodbugreport}, bug assignment~\cite{anvik2006who}, and misclassification~\cite{herzig2013misclassification}. Agent bugs complicate this line of work because the repair target may be the backend model or the harness code.

\section{Conclusion}
\label{sec:conclusion}

This paper studies Agent-Reactive (AR) bugs in agents, i.e., failures whose manifestation depends on both a specific LLM behavior and how the harness handles it. From issues across Codex, Gemini-CLI, LangChain, and CrewAI, we collect the issue reports of 255 AR bugs and analyze their symptoms, triggering LLM behaviors, and repair-target discussions by examining the bug reports and related communications. We summarized 5 symptoms, 8 triggering behaviors of AR bugs, and repair strategies or discussions for each triggering behavior from both the user and developer perspectives.
Overall, our findings shed light on reporting, reproducing, diagnosing, and repairing AR bugs, and motivate agent-specific support for test oracles, behavior-preserving reproduction, and harness-versus-model fault localization.

\clearpage

\bibliographystyle{IEEEtran}
\bibliography{references}

\end{document}